\documentclass{iopjournal}
\usepackage{amsmath,amsthm,amssymb}
\usepackage{appendix}
\newtheorem{theorem}{Theorem}
\newtheorem{lemma}[theorem]{Lemma}

\begin{document}

\articletype{Paper} %	 e.g. Paper, Letter, Topical Review...

\title{Bio-inspired learning algorithm for time series using Loewner equation}

\author{{\sc{Yusuke Kosaka Shibasaki}}$^1$$^2$\orcid{0000-0001-9491-4118}}

\affil{$^1$Institute of Natural Sciences, College of Humanities and Sciences, Nihon University,  
Setagaya, Tokyo 156-8550, Japan}

\affil{$^2$College of Art, Nihon University, Nerima, Tokyo, 176-8525, Japan}

% \affil{$^*$Author to whom any correspondence should be addressed.}

\email{shibasaki.yusuke@nihon-u.ac.jp}


\begin{abstract}
Though the relationship between the theoretical statistical physics and machine learning techniques has been a well-discussed topic, the studies on the mechanism of learning inspired by the biological system are still developing. In this study, we investigate the application methods of Loewner equation to the learning algorithm particularly focusing on its statistical-mechanical aspects. We suggest two simple methods of learning of one-dimensional time series based on the unique encoding property of the discrete Loewner evolution. The first one is the Gaussian process (GP) regression using the normality of the distribution of Loewner driving force corresponding to the curve composed from the time series. The second one is the fluctuation-dissipation relation (FDR) for the time series, which is derived from the Loewner theory, measuring the sensitivity of the nonlinear dynamics under the small perturbation. These methods were numerically tested dealing with the neuronal dynamics generated by the leaky integrate-and-fire model. In addition, we discuss the similarity between the mapping mechanism of the present method and the structure of biological information processing from a point of view of self-organization system theory. 
\end{abstract}

\section{Introduction}
Whereas the relationship between the technique of the machine learning in artificial intelligence (AI) and statistical physics is an essential issue in the current information-based science and humanities, one of the approaches that fills this gap is treating it as the problem of the complex systems [1-4]. Because the learning algorithm itself has been developed derived from the studies to model the functions of the brain (See, e.g., [5,6]), it is closely related to the nonlinear dynamics of the neural system and the question of how we treat its non-equilibrium (but robust) state as a dynamical system. As is well-known, the neural network method significantly contributes to the machine learning technique in AI enabling the computation of the nonlinear and high-dimensional problems [5-8], and this fact is also studied in statistical physics context. In the several studies, the methods in the research of non-equilibrium statistical physics (e.g., stochastic thermodynamics) have been applied to learning algorithm [9-11], where the mechanism of learning is interpreted in the physical sense. Such attempts are important not only for the development of the machine learning algorithm but also for our understanding of the physical meaning of the AI. 
As research of the general mechanisms of nonlinear (and thus non-equilibrium) phenomena in the real-world systems, the studies of Loewner equation [12-19] provide a worth discussing issue to consider the above problem. The previous studies on the physical application of the chordal Loewner equation [12] suggested that the unique encoding property of stochastic Loewner evolution (SLE) and discrete Loewner evolution [13,14], involving the chaotic dynamical system [16]. As is well-known, the chordal Loewner equation consists of two components, that is, the complex-valued curve $\gamma_{\left[0,s\right]}$ in the upper half-plane $\mathbb{H}$ and real-valued 1D function $U_s$. For the practical application, the increments $\eta_s\left(n\right)$ of the Loewner driving function $U_s$ converted from the discretized points on the curve $\gamma_{\left[0,s\right]}$ obtained via the discrete Loewner evolution is useful for the analyses of the 2D lattice systems [18] and 1D time series [19]. We refer to this dynamical variable $\eta_s\left(n\right)$ as the Loewner driving force, which has a one-to-one correspondence to the curve $\gamma_{\left[0,s\right]}$ that we analyze. For the stationarity and mixing property of the Loewner driving force [16-18], the entropy of $\eta_s\left(n\right)$ plays a central role in its formalism and the interpretation of complexity in the nonlinear dynamics [18,19]. We refer to this entropy as the Loewner entropy $S_{Loew}$, and it is expected to be an important quantity for the information analyses based on the Loewner equation. Because this Loewner equation-based information processing is enabled by the unique characteristic of conformal mapping system, the implementation of this theoretical framework to the machine learning method has a novelty and will provide some benefits to the understanding of the biological learning. 
For the above motivation, in this paper, we discuss the application method of discrete Loewner evolution to the learning algorithm. Particularly, we demonstrate that the relations obtained from the statistical-physical applications of Loewner equation (e.g., fluctuation-dissipation relation (FDR) and statistics of the Loewner driving force and entropy [18,19]) work as those help to answer this fundamental problem between machine learning and physics. As one of the simplest methods of learning, we briefly review the notions of “regression” in the machine learning [7,20]. This paper consists of the following parts. In Sec. 2.1., we present the basics of the Loewner equation, driving force, and entropy. In Sec. 2.2., as an example of the most familiar methods of leaning, the Gaussian process (GP) regression is briefly described. In Sec. 3.1., following the previous section, a possible form of the GP regression using discrete Loewner evolution is demonstrated. In Sec. 3.2., the FDR using the Loewner entropy is discussed in the context of machine learning. Subsequently, in Sec. 4., the numerical simulation of the present method is performed using time series obtained by the leaky integrate-and-fire model. Finally, in Sec. 5 and 6, the discussions and conclusion of the present study are remarked, respectively.

\section{Model}
\subsection{Loewner Equation, Driving Force, and Entropy}
Let us consider the curve $\gamma_{\left[0,s\right]}$ starting from the origin O of the upper half-plane $\mathbb{H}$. The family of the conformal map $g_s\left(z\right)=z+2s/z+O\left(z^{-2}\right)$\ (as $z\rightarrow\infty$) that transforms the region $\mathbb{H}\setminus\gamma_{\left[0,s\right]}$ to $\mathbb{H}$ is given by the following (chordal) Loewner evolution [12,14-19]: 
\begin{equation}
\frac{\partial g_s\left(z\right)}{\partial s}=\frac{2}{g_s\left(z\right)-U_s},\ \ \ \ \ \ \ \ g_0\left(z\right)=z\in\mathbb{H}.
\end{equation}
Here, $U_s$ is a real-valued 1D function called the driving function. We note that Eq. (1) becomes the standard stochastic Loewner evolution (SLE) if we choose the driving function as $U_s=\sqrt\kappa B_s$, where $B_s$ is the standard Brownian motion, i.e., Wiener process and $\kappa$ is the constant parameter that determines the phases and fractal dimension $d_f$ of the curve. The relation between the coordinate of the curve and driving function is expressed as: 
\begin{equation}
\lim_{z\rightarrow\gamma_s}{g_s\left(z\right)}=U_s.
\end{equation}
The above relation ensures the one-to-one correspondence relation of the curve and driving function. For the practical calculation of simulating the curve $\gamma_{\left[0,s\right]}$ from the driving function $U_s$, some discretizing methods of the Loewner equation (e.g., Refs. [13,14]). In this study, we employ the zipper-algorithm using the vertical slit map [14] briefly described as follows. Let us denote the curve $\gamma_{\left[0,s\right]}$ as that consists of the sequence of discretized points on $\mathbb{H}$, i.e., $\gamma_{\left[0,s\right]} =\{z_0(=0),z_1,\ldots,z_n,\ldots,z_N\} $. Using the map $h_n\left(z\right):=\sqrt{(z-\Delta U_{s_n})^2+4\Delta s_n}$ where $\Delta U_{s_n} := U_{s_n}-U_{s_{n-1}}$ and $\Delta s_n: = s_{n}-s_{n-1}$. The sequence of the increments of the driving function $\{\Delta U_{s_n}\}$ and $\{\Delta s_n\}$ corresponding to an arbitrary curve $\gamma_{\left[0,s\right]}$ is obtained from the following recursive rule [14,16-18]: 
\begin{equation}
\begin{gathered}
w_1=z_1,\ w_2=h_1\left(z_2\right),\ldots,\ w_n=h_{n-1}\circ\ h_{n-2}\circ\cdots\circ\ h_1\left(z_n\right),\ \ldots, \\
\ w_N=\ h_{N-1}\circ h_{N-2}\circ\cdots\circ h_1\left(z_N\right)
\end{gathered}
\end{equation}
where $\{\Delta U_{s_n}\}$ and $\{\Delta s_n\}$ are obtained as [14,16-18]: 
\begin{equation}
w_n=\Delta U_{s_n}+2i\sqrt{\Delta s_n}, \ \ \ \ \  i=\sqrt{-1}.
\end{equation}
For the analyses of the geometrical properties of the $\gamma_{\left[0,s\right]}$, we define the following dynamical variable:
\begin{equation}
\eta_s\left(n\right) := \frac{\Delta U_{s_n}}{\sqrt{\Delta s_n}}, \ \ \ \ \ \Delta s_n  \rightarrow0^+,
\end{equation}
which we call the Loewner driving force. For an arbitrary curve $\gamma_{[0,s]} $, the mixing property of the Loewner driving force $\eta_s\left(n\right)$, in the sense that it has an invariant probability density, has been shown using transfer operator method [18]; this property was numerically verified also with several physical systems [16,17,19]. From the uniqueness of the Loewner map and one-to-one correspondence of the curve and driving function, the Loewner driving force $\eta_s\left(n\right)$ is assumed to have full information about the development of the curve. For this reason, it is convenient we further define the entropy of the Loewner driving force as [18,19]:\footnote{In a physical point of view, if the Loewner driving function $U_s$ has a stationary distribution, this definition of the Loewner entropy is also interpreted as $S_{Loew} = k\ln\Omega(U_s)$, where $k$ is a suitable constant and $\Omega(U_s)=1/p(U_s)$ is the number of microstates of $U_s$.}
\begin{equation}
S_{Loew}:=-\ln p(\eta_s(n)).
\end{equation}
The practical usage of $S_{Loew}$ has been demonstrated in Refs. [17-19]. In Ref. [16], it was numerically indicated that $p\left(\eta_s\left(n\right)\right)$ estimated by a single time series of $\eta_s\left(n\right)$ has a distribution close to the Gaussian distribution. Furthermore, in Ref. [19], $\exp{\left(-S_{Loew}\right)}=p\left(\eta_s\left(n\right)\right)$ calculated from an ensemble of $\eta_s\left(n\right)$ has the distribution which has a good agreement with the Gaussian distribution. This Gaussian property of the Loewner driving force $\eta_s\left(n\right)$ is an important feature for considering the learning algorithm we discuss in the present study, and its theoretical reason is provided by the central limit theorem (CLT) [See, Appendix. A].

\subsection{Gaussian Process and Learning}
One of the simplest schemes to explain the term “learning” in the studies of machine learning is described as follows [7,20]. Let us consider a function $f$ that provides the relation between the output variable $y_n\in\mathbb{R}^n$ from an input variable $x_n\in\mathcal{X}^2$. 
\begin{equation}
y_n=f\left(x_n\right).
\end{equation}
If we can estimate the function $f$ from given data sets of $x$ and $y$, the prediction of the future behaviour of the system of Eq. (7) is possible, that is, 
\begin{equation}
{y_n}^\prime=f\left({x_n}^\prime\right),
\end{equation}
where, ${x_n}^\prime$ is an arbitrary input and ${y_n}^\prime$ is the predicted output. The term “regression” is used for the estimation of the function $f$, and several methods are applied. In this study, we discuss the regression method using the Gaussian process, which provides the basic idea for the present learning method using Loewner entropy $S_{Loew}$. The procedure of the GP regression is briefly summarized as the following. Let us consider the situation that the model in Eq. (7) suffers from small noise perturbation $\xi_n$ [7,20]. 
\begin{equation}
y_n=f\left(x_n\right)+\xi_n.\ 
\end{equation}
Here, $\xi_n$ denotes the white Gaussian noise with mean 0 and variance $\sigma^2$. From Eq. (9), we obtain the following relation.
\begin{equation}
y_n-f\left(x_n\right)=\xi_n\sim\mathcal{N}\left(0,\ \sigma^2\right),
\end{equation}
where $\mathcal{N}$ denotes the normal distribution. From Eq. (10), the likelihood function between $x_n$ and $y_n$ is expressed as:
\begin{equation}
p\left(y\middle| X\right)=\prod_{n=1}^{N}\mathcal{N}\left(y_n|f(x_n),\ \sigma^2\right).
\end{equation}
Using Bayne’s rule, the posterior distribution is obtained as [20]:
\begin{equation}
dp_n\left(f(x_n)\right)\propto p\left(y\middle| X\right)dp_0\left(f(x_0)\right)=\prod_{n=1}^{N}\mathcal{N}\left(y_n|f(x_n),\ \sigma^2\right)dp_0\left(f(x_0)\right).
\end{equation}
It is found that the posterior distribution also has a form of a Gaussian process, where its mean and covariance are expressed by the parameters involving kernel function [7,20]. From the probabilistic relation in Eq. (12), the estimation of the behaviours of the output variable from the new (unknown) input becomes possible from the relation between the known dataset $x_n$ and $y_n$. The above GP regression procedure is one of the representative methods of the machine learning. 

\section{Results}
\subsection{Gaussian Process Regression using Loewner Equation}
In this subsection, we shall introduce an application method of Loewner equation to the above-described GP regression method for learning. As we see, the essence of the GP regression is that, for arbitrary input and output variables, there exists a Gaussian process that provides the likelihood function and posterior distribution, which also are the Gaussian processes. This property is also found in the theoretical scheme of discrete Loewner equation as the one-to-one correspondence relationship between an arbitrary curve $\gamma_{\left[0,s\right]}$ and driving function $U_s$, whose increment $\eta_s(n)$ forms a Gaussian distribution. In the followings, we show a method of GP regression using Loewner equation, which is intrinsically equivalent to that described in Sec. 2.2. Let us consider the problem of how we predict the future behaviour of the time series ${x_n}$ from that of its history. To implement 1D time series into the theoretical scheme of Loewner equation, we compose the curve $\gamma_{\left[0,s\right]}$ from the time series as: 
\begin{equation}
\gamma_{\left[0,s\right]}:= \{z_0(=0),z_1=x_1+i\tau, z_2=x_2+2i\tau\ldots,z_n=x_n+ni\tau,\ldots,z_N=x_N+Ni\tau\}.  
\end{equation}
Here, $\tau$ is a sufficiently small constant. From Eqs. (2)-(5), we observe that the relation between $z_{n+1}$ and $z_n$ is expressed as the following:
\begin{equation}
z_{n+1}=g_{n+1}^{-1}(g_n\left(z_n\right)+c(\tau)\eta_s(n)).
\end{equation}
Here, $g_n$ denotes the map $g_s$ corresponding to the time point $s_n$ and $c(\tau)$ is a $\tau$-dependent constant. The relation in Eq. (14) is immediately transformed as:
\begin{equation}
g_{n+1}\left(z_{n+1}\right)=g_n\left(z_n\right)+c(\tau)\eta_s(n).
\end{equation}
Because of its mixing property, the Loewner driving force $\eta_s\left(n\right)$ in the above relation works in a similar manner as the white Gaussian noise although it may contain some autocorrelation depending on the dynamical property [16,18]. From the CLT for the mixing dynamical system, $p\left(\eta_s\left(n\right)\right)$ forms a Gaussian distribution denoted as $\mathcal{N}\left(\mu,\sigma^2\right)$, where $\mu$ and $\sigma^2$ are its mean and variance, respectively. To connect the theory of Loewner equation and the GP regression, we assume that the relation in Eq. (15) is analogous to that in Eq. (9). For the above arguments, we note that the conditional probability $p\left(g_{n+1}\left(z_{n+1}\right)\middle|\ g_n\left(z_n\right)\right)$ is calculated as:
\begin{equation}
p\left(g_{n+1}\left(z_{n+1}\right)\middle| g_n\left(z_n\right)\right)=\left|J\right|p\left(U_{n+1}\middle| U_n\right)=\left|J\right|p\left(\eta_s\left(n\right)\right)\sim C(n)\mathcal{N}\left(\mu(\tau),\sigma^2\right(\tau))
\end{equation}
where, $C\left(n\right):=\left|J\right|$ is the Jacobian of the conformal map $g_n$ expressed as:
\begin{equation}
C\left(n\right):=\left|J\right|=\left|\frac{dg_n\left(z_n\right)}{dz}\right|=\left|1-\frac{2s}{{z_n}^2}+O\left({z_n}^{-3}\right)\right|.
\end{equation}
Denoting $Z=\{{z_k}\}_{k=0}^n$ and using Eqs. (16) and (17), the likelihood function is obtained as:
\begin{equation}
p\left(z_{n+1}\middle| Z\right)=\prod_{n=1}^{N}{C(n)\mathcal{N}\left(\mu(\tau),\ \sigma^2(\tau)\right)}.
\end{equation}
Using Bayne’s rule, the posterior distribution is obtained as the following [20]: 
\begin{equation}
dp_{n+1}\left(z_{n+1}\right)\propto p\left(z_{n+1}\middle| Z\right)dp_0\left(z_0\right)=\prod_{n=1}^{N}{C(n)\mathcal{N}\left(\mu(\tau),\ \sigma^2(\tau)\right)}dp_0\left(z_0\right).
\end{equation}
It is found that Eq. (19) has a similar form to Eq. (12) for a sufficiently large $n$ because $C(n)\rightarrow \rm Const.$ as $z_n$ becomes large in Eq. (17).\footnote{In the numerical simulation of this study, we fixed the term $C(n)$ as $C(n)=1.0$, and $\mu(\tau)$ is adjusted to that of the single input time series $x_n$ as presented in Sec.4.} Therefore, the present approach is intrinsically equivalent to GP regression in a usual sense. It should be mentioned that the term $c(\tau)$ is theoretically determined by the parameters and amplitudes of time series that we choose; however, it is also be interpreted as the noise strength parameter of the GP regression. Therefore, in the practical applications, it is treated as an arbitrary constant parameter, and we replace it as $c(\tau)\rightarrow\beta$, which denotes the noise strength in Eq. (15). In this manner, the variance $\sigma^2$ is determined by external parameter $\beta$ similarly to the usual GP regression although the qualitative behaviour of the posterior distribution in Eq. (19) is singly determined by the dynamics of the Loewner driving force $\eta_s\left(n\right)$. In addition, to observe the use of Loewner entropy defined in Eq. (6), we introduce log-likelihood function expressed as: 
\begin{equation}
\begin{split}
\ln{p\left(z_{n+1}\middle| Z\right)}&=\ln{\prod_{n=1}^{N}C\left(n\right)\mathcal{N}\left(\mu,\ \sigma^2\right)}\\
&=-\sum_{n=1}^{N}{C\left(n\right)S_{Loew}}.
\end{split}
\end{equation}
From Eq. (20), we found that the minimization of Loewner entropy $S_{Loew}$ corresponds to the maximization of the log-likelihood function. The above relation expresses the maximum likelihood method in GP regression is replaced by the theoretical framework of the Loewner equation.

\subsection{Statistical-Mechanical Approach for Learning}
In the previous subsection, we observed the implementation of GP regression for time series using discrete Loewner evolution. The method is based on the mixing property of Loewner driving force $\eta_s\left(n\right)$ and Gaussian property of its probability density function $p\left(\eta_s\left(n\right)\right)=\exp{\left(-S_{Loew}\right)}$ (See, Appendix. A). As we showed, this approach is useful for the estimation of the posterior distribution of the output variable from the prior distribution of input variable. For the time series analyses of nonlinear systems; however, the effect of small perturbation to the output of the system is an important issue to be clarified. In the context of studies of nonlinear science, it is often measured by Lyapunov exponents, Kolmogorov-Sinai entropy, etc. In this view, it is worth noting that the previous studies on Loewner equation suggested that (nonlinear) fluctuation-dissipation relation (FDR) that is applicable for the 1D time series. For this reason, we here describe the application method of FDR using Loewner theory and its meaning of “learning”. Let us consider the following relation modified from Eq. (9).
 \begin{equation}
 dy_n=df(x_n)+dc_{n^\prime}\ \ \ \ \ \ n>n^{\prime}.
 \end{equation}
Here, $x_n$ and $y_n$ are input and output variables, and $c_n$ is a small perturbation to the system $f$. The indices $n^\prime $ and $n$ denote the time point in which the perturbation occurs, and response is observed, respectively. For the above settings, the response function $R\left(n,n^\prime\right)$ is defined as the following [21,22]: 
\begin{equation}
R\left(n,n^\prime\right):=\frac{d\langle y_n\rangle}{dc_{n^\prime}}=\frac{d\langle f(x_n)\rangle}{dc_{n^\prime}}+1.
\end{equation}
In the r.h.s. of Eq. (22), we used the relation in Eq. (21). Defining $V_s(n):=\sum_{k=1}^n \eta_s(k)$, the time conversion method using Loewner time $\{s_n\}$ leads to the following expression of the response function [16,23,24]:\footnote{In this formalism, we assume the one-to-one correspondence of the curve and driving force and replace the ensemble average of the nonlinear response function into that of the equilibrium distribution of $S_{Loew}$. For the application to the Langevin-type equation, in this study, the diffusion coefficient $D$ is used.}  
\begin{equation}
R\left(n,n^\prime\right)=\frac{1}{2D}{\langle V_s(n)\Delta V_s(n^\prime)\rangle}_{eq}.
\end{equation}
Using Eqs. (6), (22) and (23), the ensemble average of the output variable $y_n$ is calculated as the following:
\begin{equation}
d\langle y_n\rangle=dc_{n^\prime}\int V_s(n)\eta_s(n^\prime)\exp(S_{Loew})d\eta_s.
\end{equation}
The relation in Eq. (24) expresses the nonlinear variation of the time series at time point $n$, when a small perturbation occurs at $n^{\prime}$. The prediction of the behavior of the system is based on the information of the history of the time series encoded into the Loewner driving force $\eta_s\left(n\right)$. In this sense, in the present scheme, the ‘learning’ is interpreted as the action of the time-dependent mapping of the Loewner equation. We note that the nonlinear response is dependent on the time distance $\left|n-n^{\prime}\right|$; therefore, the precision of the prediction is determined by the length of the time series that we observe. In the next section, we show the results of the numerical simulation of the leaning algorithm we presented in this section. 

\section{Numerical Simulation}
\subsection{Learning the time series of neuronal dynamics}
The numerical simulations have been performed to verify the results of the algorithm introduced in Sec. 3.1. and Sec. 3.2. The time series used in this study is the neuronal dynamics generated by the leaky integrate-and-fire model expressed as [19,25,26]:
\begin{equation}
\frac{dv\left(t\right)}{dt}=-\left(v-v_R\right)+m+c\left(t\right)+\sqrt{2D}\xi\left(t\right).
\end{equation}
Here, $v\left(t\right)$ is the voltage across the membrane, and m and $c\left(t\right)$ are constant and time-dependent input current, respectively. In this study, we choose $c\left(t\right)$ as $c\left(t\right)=Acos(\omega\ t)$. The term $\xi\left(t\right)$ is the white Gaussian noise with mean $0$ and variance $1.0$. The diffusion coefficient $D$ expresses the strength of the noise. The simulations were performed such that the voltage $v\left(t\right)$ resets to $v_R$ when it reaches to the threshold value $v_{th}$. We simulated the neuronal dynamics expressed in Eq. (25) using the Euler method with the following fixed parameter: $v_{th}=1.0,\ \ v_R=0.0,\ \ m=2.0,\ \ \omega=10.0,\ D=0.1$ and $\ \tau=0.001$. The length of the time series is set as $N=10000$, and the statistics of the time series is calculated over $500$ realizations ($M=500$). Figure 1 shows the results of GP regression using Loewner equation. In the upper figure, the blue solid line represents the time series of the voltage $v\left(t\right)$, and the area filled with shade expresses the range of $2\beta\sigma$, calculated from the distribution of the Loewner driving force $\eta_s(n)$ with noise strength $\beta=0.25$. The time series of the Loewner driving force $\eta_s(n)$ corresponding to each time series of $v\left(t\right)$ are plotted in the lower figures, and the standard deviation $\sigma$ was calculated for every time point $n$. It was observed that the width of $2\beta\sigma$ is relatively linear if the nonlinearity of the dynamics of $v\left(t\right)$ determined by $A$ is small (See, Fig. 1(a) for $A=0.05$ and 1(b) for $A=10.0$.) In addition, the width of $2\sigma$ is $\tau$-dependent, and a small value of $\tau$ results in a large $\sigma$. This tendency is interpreted as that related to the precision of the prediction (See, Sec. 4.1. and Appendix.B). Figure 2 shows the histograms of the term $\exp{\left(-S_{Loew}\right)}$ obtained in the simulations (1(a) for $A=0.05$ and 2(b) for $A=10.0$). For these plots, the Gaussian property of the term $\exp{\left(-S_{Loew}\right)}$ was numerically confirmed, and this result is important for the statistical-mechanical formalism of FDR in Eqs. (23) and (24). Figure 3 shows the results of FDR method presented in Sec. 3.2. It is assumed that the small perturbation is added in the first time point ($n=1$) of the time series. For the calculation of the variation of $v\left(t\right)$ under the small perturbation, we apply the following relation derived from Eq. (24).
\begin{equation}
\langle \Delta v(t)\rangle = \frac{1}{Mn}\sum_{h=1}^M \sum_{k=1}^n \left(\frac{\varepsilon}{2D}\right)V_s(k, h)\eta_s(n^\prime=1, h)\exp{\left(-S_{Loew}\right)}.
\end{equation}
Here, $\varepsilon$ denotes the parameter that determines the strength of the perturbation. In this study, we set it as $\varepsilon=0.0001$. The plots of Fig. 3 (a) and 3(b) show the results of the simulation for $A=0.05$ and $A=10.0$, respectively. The blue solid line represents the time series of the voltage $v\left(t\right)$, and the area filled with shade expresses the range of $2\sigma$ of $v\left(t\right)$ calculated by Eq. (26). The time series of the Loewner driving force $\eta_s\left(n\right)$ corresponding to each time series of $v\left(t\right)$ are the same as those plotted in the lower figures in Fig. 1, and the Loewner entropy $S_{Loew}$ was calculated using the sample entropy method [27] (in the same manner as Fig. 2). By comparing Figs. 3(a) and (b), the nonlinearity-dependence of the precision of the prediction is also found in this method, and it was confirmed that the uncertainty of the prediction is highly dependent on $\varepsilon$. These results suggest that the FDR method measures the sensitivity of the initial condition in a form different from those based on the conventional method such as the Lyapunov exponent.  \footnote{Though the theoretical model of the neuronal dynamics $v(n)$ has the threshold values, in this study, the predictions were performed solely depending on the numerically observed time series data. In other words, we assume the experimental and/or real-world data-based situations in which we do not have knowledge about the model existing behind the observed time series.} 

\begin{figure}[htbp]
\centerline{\includegraphics[width=13cm]{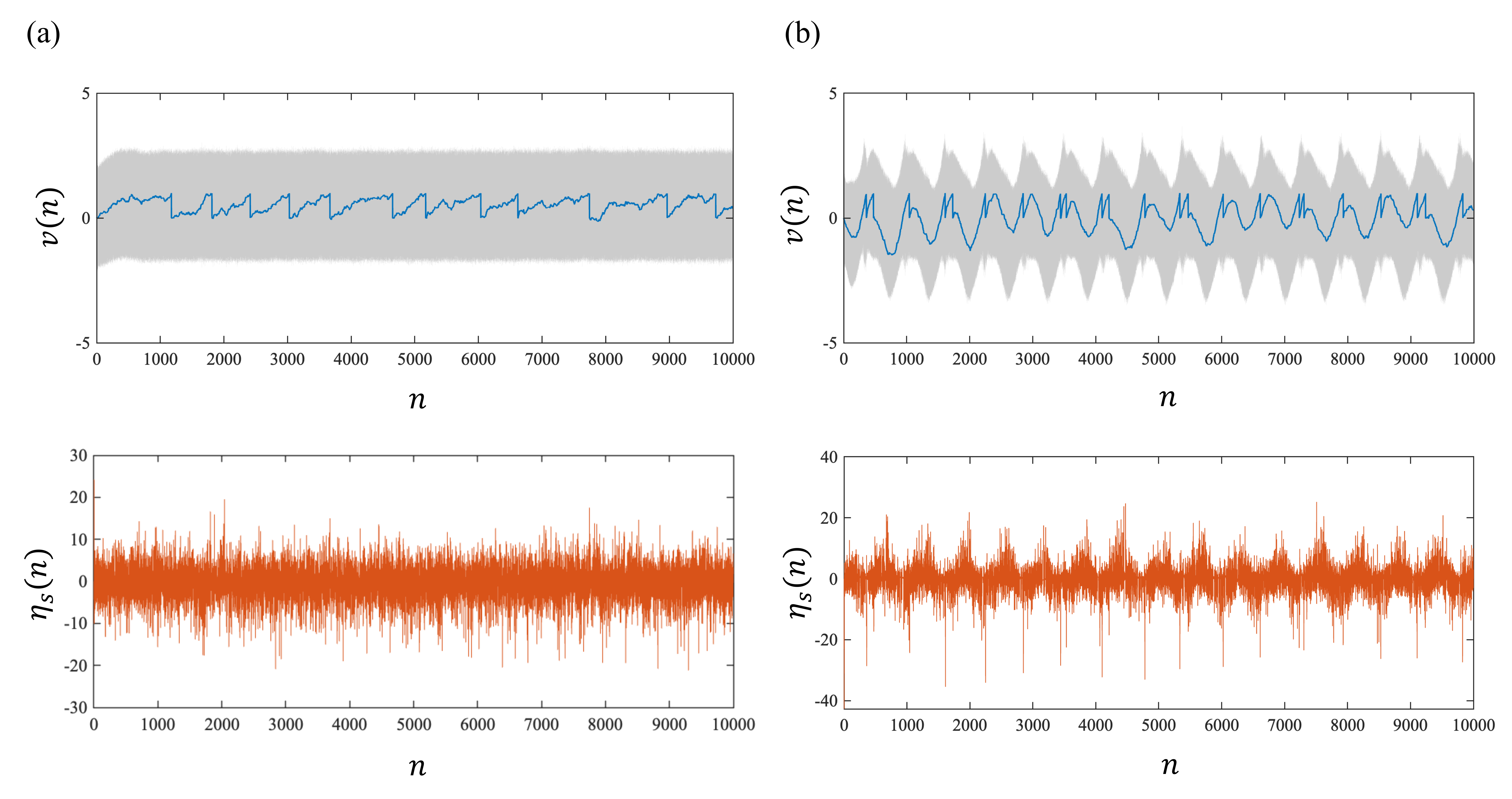}}
\vspace*{8pt}
\caption{GP regression using Loewner equation of time series of neuronal dynamics generated by the leaky integrate-and-fire model. (a) Plot of neuronal dynamics $v(n)$ with $A=0.05$ (upper) and that of the Loewner driving force $\eta_s (n)$ (lower). (b) Plot of neuronal dynamics $v(n)$ with $A=10.0$ (upper) and that of the Loewner driving force $\eta_s (n)$ (lower). For both plots, the blue plot in the upper figure represents $v(n)$ and grey shade area represents the range of $\overline{\mu}\pm2\beta\sigma$ calculated from 500 realizations of $\eta_s(n)$ ($\overline{\mu}$ represents the time average of the plotted time series of $v(n)$). The other parameters in the leaky integrate-and-fire model in Eq. (25) are noted in the main text.}
\end{figure}

\begin{figure}[htbp]
\centerline{\includegraphics[width=12cm]{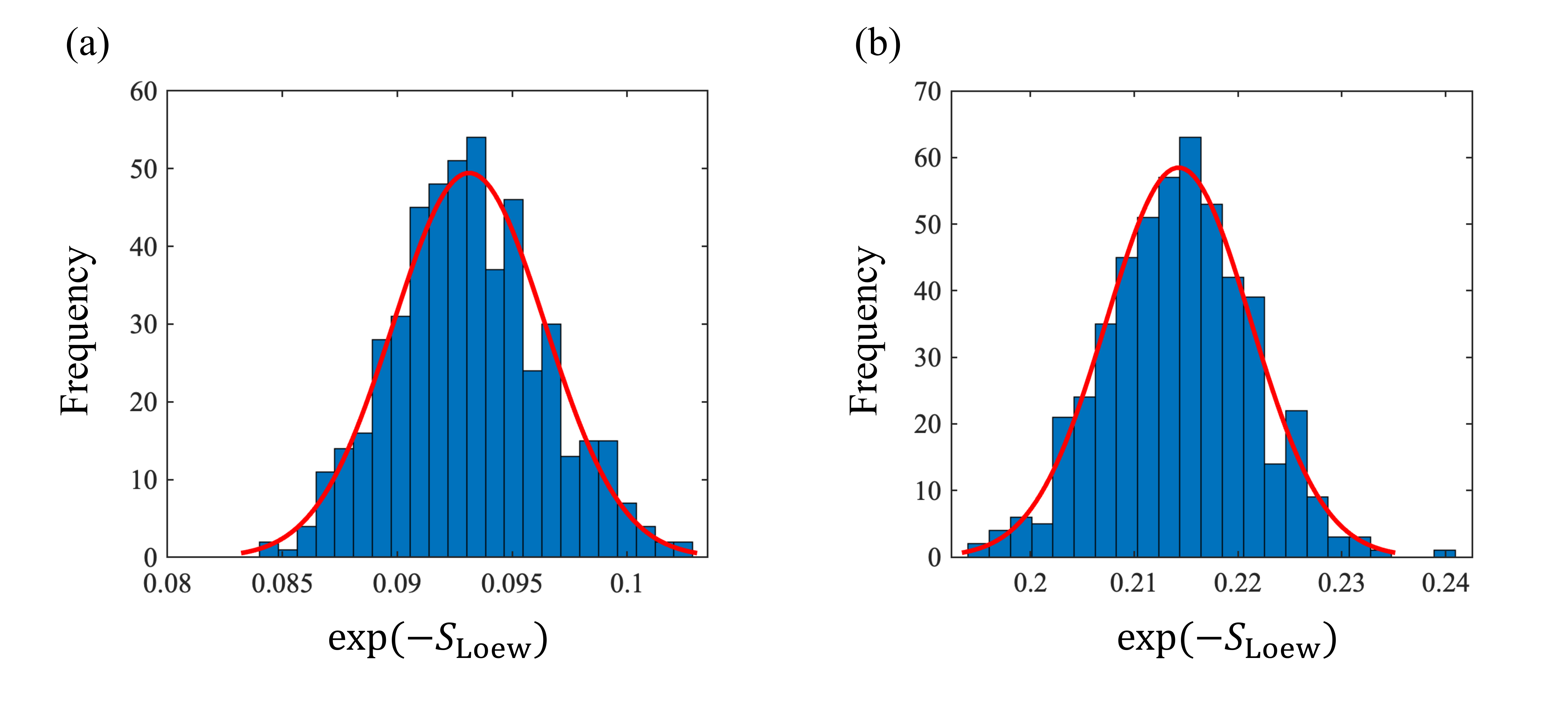}}
\vspace*{8pt}
\caption{The distribution of $\exp{\left(-S_{Loew}\right)}$. (a) Histogram of $\exp{\left(-S_{Loew}\right)}$ calculated from neuronal dynamics $v(n)$ with $A=0.05$. The red plot shows the fitting of Gaussian distribution with $\mu(\tau)=0.0931$ (mean) and $\sigma(\tau)=0.0033$ (S.D.). (b) Histogram of $\exp{\left(-S_{Loew}\right)}$ calculated from neuronal dynamics $v(n)$ with $A=10.0$. The red plot shows the fitting of Gaussian distribution with $\mu(\tau)=0.2143$ (mean) and $\sigma(\tau)=0.0070$ (S.D.). }
\end{figure}

\begin{figure}[htbp]
\centerline{\includegraphics[width=10cm]{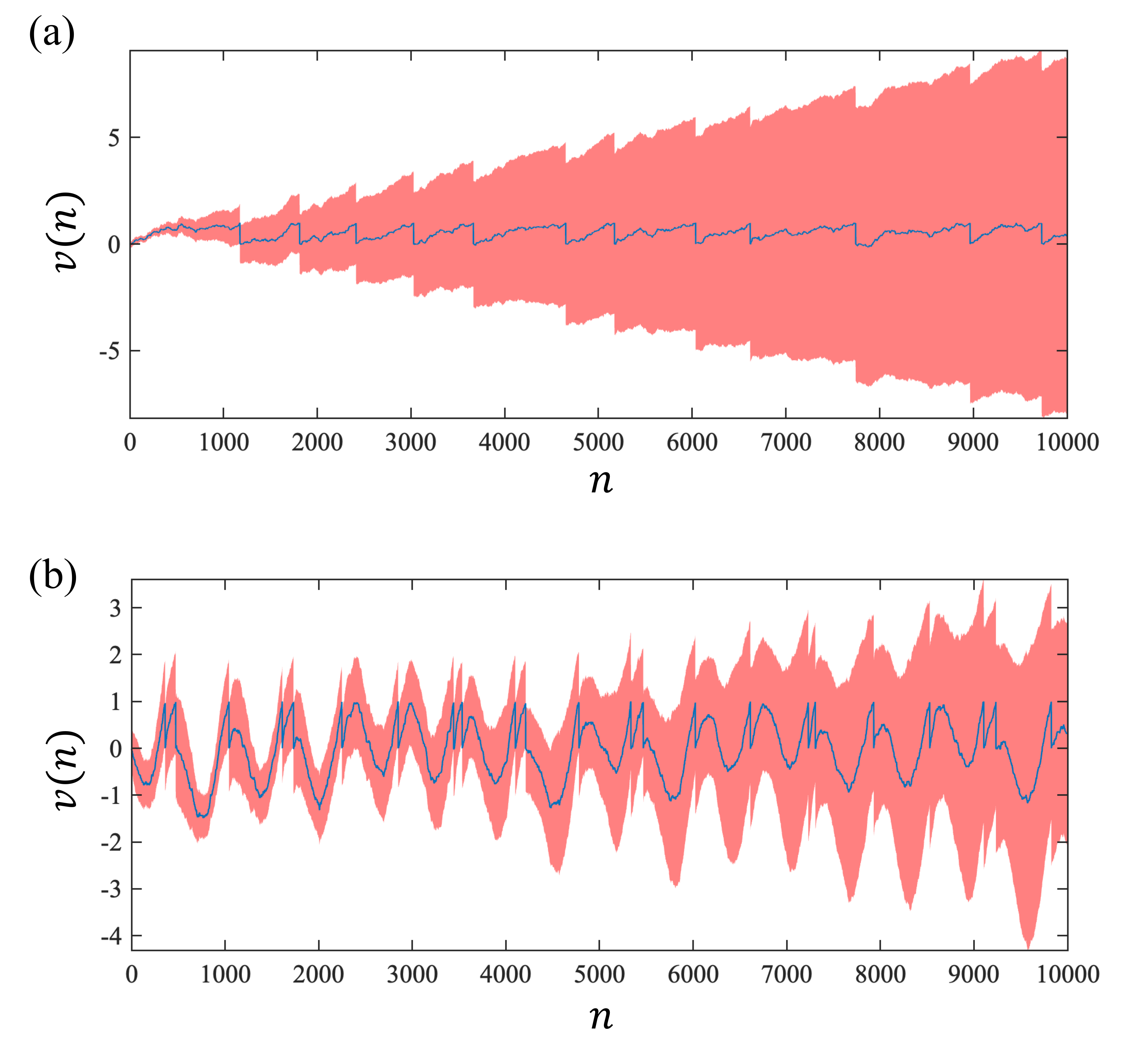}}
\vspace*{8pt}
\caption{FDR method of the prediction of time series of neuronal dynamics generated by the leaky integrate-and-fire model under a small perturbation. (a) Plot of neuronal dynamics $v(n)$ with $A=0.05$ (blue plot) and the range of $v(n)\pm2\sigma$ (red shaded area). (b) Plot of neuronal dynamics $v(n)$ with $A=10.0$ (blue plot) and the range of $v(n)\pm2\sigma$ (red shaded area). The strength of the perturbation is set as $\epsilon=0.0001$. The other parameters in the leaky integrate-and-fire model in Eq. (25) are noted in the main text and the same as Fig.1. }
\end{figure}

\subsection{$\tau$-dependence of the $\sigma(\tau)$}
As we noted in the previous sections, the learning algorithm based on the calculation of $\sigma(\tau)$ of the Loewner driving force $\eta_s(n)$ is dependent on the input variable $\tau$. In this subsection, we investigate this effect in a more detailed manner for the choice of $\tau$ in the algorithm. Figure 4 represents the numerical results on $\tau$-dependence of the $\sigma(\tau)$, showing the log-log plot of $\tau$ and $\overline{\sigma(\tau)}$ (i.e., the time average of $\sigma(\tau)$) for the conditions $A=0.05$ and $A=10.0$. We observe that two different scalings of $\overline{\sigma(\tau)}$ with respect to $\tau$; $\overline{\sigma(\tau)}\sim\tau^{-0.5}$ in the regime close to $\tau=0.0$ and $\overline{\sigma(\tau)}\sim\tau^{-0.25}$ in the regime close to $\tau=1.0$, respectively. It was found that this scaling becomes stable from the regime less than $\tau\sim10^{-2}$. This result provides the reasoning for the precise choice of $\tau$ in the practical situations, and the mathematical explanation for this result is also provided on Appendix B.
 
\begin{figure}[htbp]
\centerline{\includegraphics[width=8cm]{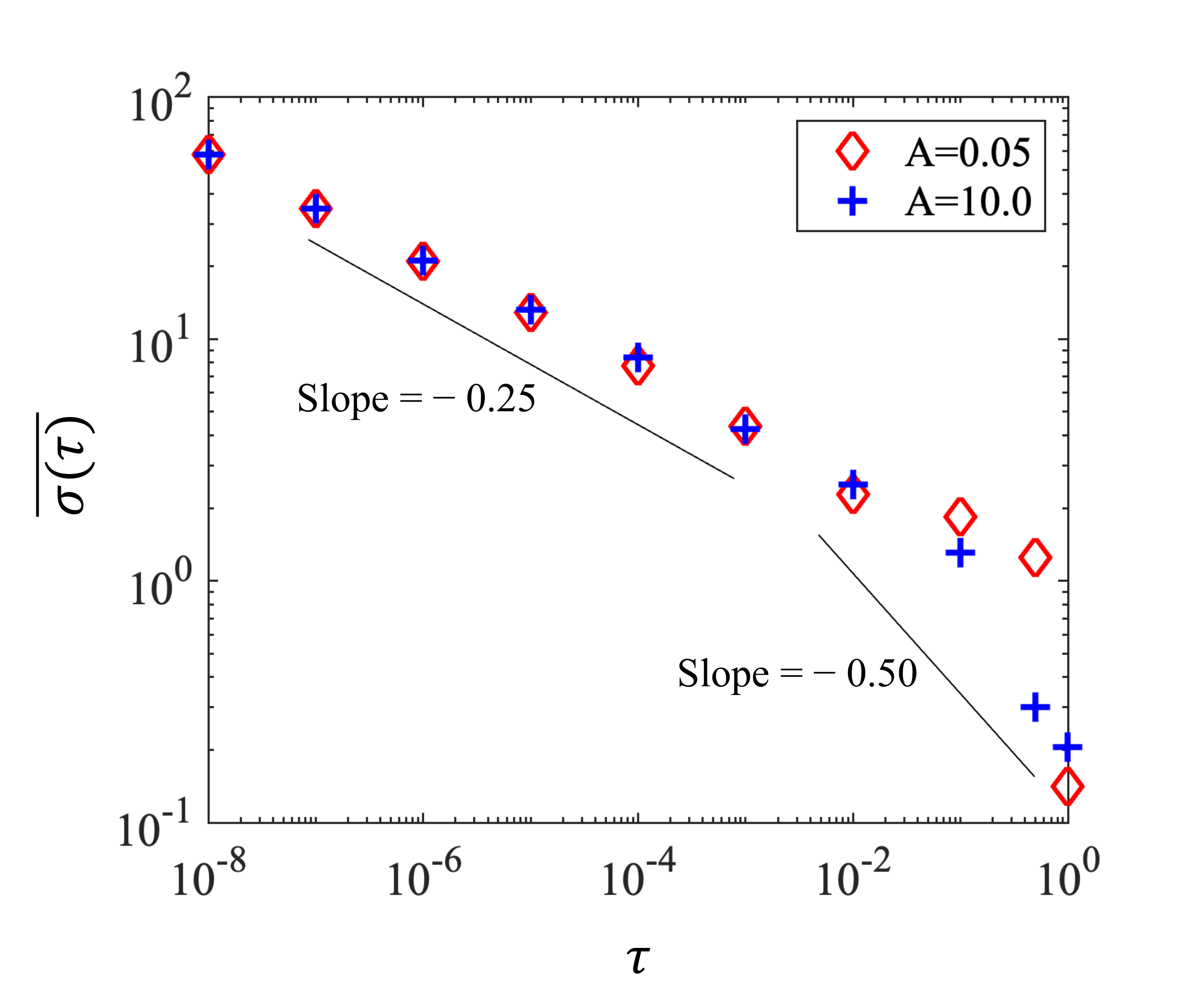}}
\vspace*{8pt}
\caption{$\tau$-dependence of $\sigma(\tau)$. Log-log plot of $\tau$ and $\overline{\sigma(\tau)}$ for the conditions $A=0.05$ and $A=10.0$. The slopes in the figure indicates that of $\overline{\sigma(\tau)}\sim\tau^{-0.5}$ and $\overline{\sigma(\tau)}\sim\tau^{-0.25}$, respectively. A mathematical explanation for the observed scalings is provided in Appendix. B.}
\end{figure}

\section{Discussion}
In this paper, the learning algorithm using Loewner equation is investigated particularly focusing on the utility of Loewner entropy $S_{Loew}$. The procedure of the method is based on the state-dependent iteration of the conformal map expressed in Eq. (3). In Sec. 3.1, we presented the GP regression method based on the normality of the probability distribution of the Loewner driving force, which is numerically verified with the leaky integrate-and-fire model as shown in Fig.1. This method is interpreted as a quantitative approach of learning implemented using Gaussian distribution. Contrarily, the FDR method presented in Sec. 3.2 is derived from the concept from the statistical mechanics, and it measures the qualitative behavior of the time series after the small perturbation as numerically shown in Fig.3. In this sense, the present method is similar to that of neural network method. For example, $L$-layer neural network is expressed by the function $f_w\left(X\right)$ expressed as [1]:
\begin{equation}
f_w\left(X\right)=g^{\left(L\right)}\left(W^{\left(L\right)}\ldots g^{\left(2\right)}\left(W^{\left(2\right)}g^{\left(1\right)}\left(W^{\left(1\right)}X\right)\right)\right). 
\end{equation}
Here, $W^{\left(n\right)} (n=1,2,\ldots,L)$ denotes the matrices of weights and $g^{\left(n\right)}$ is a $n$-dependent function. The fitting to the nonlinearity of the data is adjusted by the state-dependent map $g^{\left(n\right)}$ and the weight matrices $W^{\left(n\right)}$. In this approach, the number of the mapping $L$ means the depth of the neural network, and the algorithm with large depth called the deep learning. Comparing Eq. (27) with discrete Loewner evolution in Eq. (3), we observe a certain similarity between them in the sense that state-dependence of the iteration of the map; however, we notice that the discrete Loewner evolution has a more complicated structure, that is, the data size (curve length)-dependence of the iteration of the map. The relation to the biological system theory and the present theoretical framework is discussed from the viewpoint of autopoietic theory [28], whose developmental (iteration) rule is expressed as [29]:
\begin{equation}
\{{\rm boundary}_{n+1},\ {\rm state}_{n+1}\}=F_{<{\rm boundary}_n,\ {\rm state}_n>}\left({{\rm boundary}_n,\ {\rm state}_n}\right).\ 
\end{equation}
In the present method, it is natural to interpret the boundary corresponds to the curve $({\mathrm{boundary}}_n=\gamma_{\left[0,s\right]})$, and the state corresponds to $x\left(t\right)$ and $\eta_s\left(n\right)$ although $\eta_s\left(n\right)$ is a hidden variable. The development rule $F_{<{\mathrm{boundary}}_n,\ {\mathrm{state}}_n>}$ corresponds to the function expressed in Eq. (14), which is a key component governing the time evolution of $x\left(t\right)$. For this reason, it is found that the present learning algorithm is thought to be “biologically inspired” in the sense of the autopoietic theory, which is structurally different from the deep neural network theory. Therefore, its relation to the real cognitive and learning mechanism of the human brain should be further investigated. 
  In addition, we remark the advantage of the computational cost of the present algorithm. For given time series that consist of $N$ data points, the zipper algorithm to compute the corresponding Loewner driving force $\eta_s\left(n\right)$ takes a calculation time of (less than) $O\left(N^2\right)$ [14]. This is faster than the usual GP regression which takes the calculation time of $O\left(N^3\right)$ [30] to calculate the covariance function from kernel function. This also means that the present method shortens the procedure of the calculation of the required GP from a kernel function, and instead it is uniquely determined by the history of the single orbit of the time series as the Loewner driving force $\eta_s\left(n\right)$. As mentioned above, this is possible by the characteristic of the iteration (developmental) rule of the discrete Loewner evolution in Eq. (3), which is reasonable in a biological sense as described in Eq. (28).

\section{Conclusions}
In conclusion, we proposed two methods of the learning algorithm using Loewner equation. The first one is the GP regression using the Gaussian property of the Loewner driving force $\eta_s\left(n\right)$ directly. The second one is based on the FDR in the statistical-mechanical sense that predicts the future behaviour of the time series. The suggested methods of learning were numerically tested using the neuronal dynamics generated by the leaky integrate-and-fire model. For the specific parameters, the validity of both methods was confirmed. We observed that the uncertainty of the prediction increases as $n$ become large in the FDR method, while the range of the uncertainty is stationary in the GP regression method. Both methods showed the predictability of the time series which is noise-strength and perturbation-strength dependent. In addition, the biologically inspired structure of the learning algorithm using discrete Loewner evolution is discussed in terms of autopoietic theory, suggesting that the similarity and difference between the present method and neural network method. Further studies will help to construct a more detailed and generalized form of conformal mapping-based approach of the machine learning, while providing a novel understanding of the real learning mechanism of the brain in the biological systems. 

%
% Each of the commands below will create an unnumbered section with the appropriate heading.
% Remove any sections that are not relevant for your article.
% All sections except suppdata will be removed if the [anonymous] option is used.
% See iopjournal-guidelines.pdf for more information.
%

\ack{The main part of this study was performed at the Institute of Natural Sciences, College of Humanities and Sciences, Nihon University. I would like to thank the courtesy of the Institute for having maintained the support of this research. The result of this paper has been presented at Statphys29, Florence, Italy and I would like to acknowledge the conference committee for giving the opportunity to present this research. }

\funding{The author did not receive any funding for conducting this research.} 

\roles{The author conducted research and wrote the article.} 

\data{The data that support the findings of this study are available from the corresponding author upon reasonable request.}
% For more information on IOP Publishing's research data policy see: https://publishingsupport.iopscience.iop.org/questions/research-data/

% \suppdata{Sample text inserted for demonstration.}

%\appendix
\renewcommand{\theequation}{A.\arabic{equation}} 
\setcounter{equation}{0} 
\section*{Appendix A. Gaussian Property of Loewner Driving Force}
The Gaussian property of the probability distribution of the Loewner driving force $\eta_s\left(n\right)$ is numerically reported in Refs. [16,19]. We here note mathematical evidence of this Gaussian property using the mixing property of the transformation in Eq. (3), which is discussed in Ref. [18]. Let us denote the initial condition of the state-dependent dynamical system in Eq. (3) as:
\begin{equation}
\gamma_{\left[0,n\right]}:=\{z_0,z_1,…,z_n\}\in\mathbb{H}.
\end{equation}
The probability distribution of $w_n$ after $k$ iterations of the map $h_n$ is expressed using the transfer operator $\mathcal{L}_n$, that is,
\begin{equation}
p_n\left(w\right)=\mathcal{L}_n^{k}p_0\left(z\right).\ 
\end{equation}
Here, the transfer operator $\mathcal{L}_n$ is defined as that satisfying the following [31]:
\begin{equation}
\mathcal{L}_np\left(w\right)=\sum_{z\in h^{-1}\left(w\right)}{\left|h_n'\left(\chi_r\left(w\right)\right)\right|^{-1}p\left(\chi_r\left(w\right)\right)}. 
\end{equation}
Note that here the sum is taken over two preimages $\chi_r\left(w\right)$ of $h_n\left(z\right)$, which is expressed as:
\begin{equation}
\chi_r\left(w\right):=h_{r=1,2}^{-1}=\Delta U_{s_n}\pm\sqrt{w^2-4\Delta s_n}
\end{equation}
and we obtain the following:
\begin{equation}
\left|h_n'\left(\chi_r\left(w\right)\right)\right|^{-1}=\left|\frac{\pm\sqrt{w^2-4\Delta s_n}}{w}\right|^{-1}.
\end{equation}
\begin{theorem}
The transformation in Eq. (3) to the curve $\gamma_{\left[0,s\right]}$ composed from the time series $x_n$ by Eq. (13) is topologically mixing, in the sense that,
\begin{equation}
\left|p_n\left(w\right)-{p_0}^*\left(z\right)\right|\ \sim\exp{\left[-K\left(n\right)\right]},\ \ and\ \ \ \  K\left(n\right)\rightarrow\infty\ \ \ \ as\ \ \ n\rightarrow\infty.
\end{equation}
where ${p_0}^*\left(z\right)$ is the invariant probability density and $K\left(n\right)$ is a function of $n$. 
\end{theorem}
The proof of the Theorem is the same as that in Ref. [18]. Accordingly, from the independence of the real and imaginary coordinates of $w_n$, the variable $\eta_s\left(n\right):=\Delta U_{s_n}/\sqrt{\Delta s_n}$ is also mixing and the exponential decay of the autocorrelation of $\eta_s\left(n\right)$ is observed if the real part of the initial condition $\gamma_{\left[0,s\right]}$ is not $n$-dependent. The probability density function of variable $\eta_s\left(n\right)$ is obtained as $p(\eta_s(n))=\ \exp{\left(-S_{Loew}\right)}$ from the definition of Loewner entropy $S_{Loew}$ in Eq. (6), and the probability of $\ \exp{\left(-S_{Loew}\right)}$ calculated from $M$ statistical ensemble at time $n$ is expressed as:
\begin{equation}
\begin{split}
\rm{E}\left[\it{p}(\eta_s\left(n\right))\right]&=\langle \exp{(-S_{Loew})} \rangle\\
&=\frac{1}{M}\sum_{\eta_s\left(n\right)\in\mathcal{W}}\exp{\left(-S_{Loew}\right)}.
\end{split}
\end{equation}
where $\mathcal{W}$ denotes the set of $\eta_s\left(n\right)$. From the independence of $\exp{\left(-S_{Loew}\right)}$, the CLT leads to the following relation:
\begin{equation}
\exp{\left(-S_{Loew}\right)}\sim\ \mathcal{N}\left(\mu\left(\tau\right),{\sigma\left(\tau\right)}^2\right).\ 
\end{equation}
Here, $\mu\left(\tau\right)$ and $\sigma\left(\tau\right)^2$ denote the mean and variance, and they are $\tau$-dependent [18]. The relation in Eq. (A8) is numerically verified also in Ref. [19]. 

%\appendix
\renewcommand{\theequation}{B.\arabic{equation}} 
\setcounter{equation}{0} 
\section*{Appendix B. Convergence of Time Increments of Loewner Driving Force}
In the main text, we assume the convergence of the time increments of the Loewner driving function $\Delta s_n \rightarrow0^+$ in the limit of $\tau\rightarrow0^+$. One of the methods to approximate this convergence is provided by the backward Loewner evolution [23,24,33]. Assuming the (time-) symmetric property of the reconstructed Loewner driving function $V_s$, the probability distribution of the time evolution of the tip $\{\rm Re\gamma_s,\ \rm Im\gamma_s\}\in\mathbb{H}$ of the curve $\gamma_{\left[0,s\right]}$ is the same as those of $\{x_n,\ t_n\}\in \mathbb{R}^2$ governed by the following two-dimensional differential equation.
\begin{align}
\frac{dx_n}{ds}&=-\frac{2x_n}{{x_n}^2+{t_n}^2}-\frac{dV_s}{ds},\\
\frac{dt_n}{ds}&=\frac{2t_n}{{x_n}^2+{t_n}^2}.
\end{align}
To investigate the relationship between the linearized time $\{t_n\}$ and nonlinear time $\{s_n\}$, in the above equations, we regard $\rm Im\gamma_s \rightarrow {\it t_n}$ [23,24]. Applying the discretization based on the Euler scheme, Eqs. (B.1) and (B.2) are rewritten as:
\begin{align}
x_{n+1}&=x_n-\Delta s_n\frac{2x_n}{{x_n}^2+{t_n}^2}-\sqrt{\Delta s_n}\eta_s(n), \\ 
t_{n+1}-t_n&=\tau=\Delta s_n\frac{2t_n}{{x_n}^2+{t_n}^2}. 
\end{align}
Because $\Delta s_n$ scales as $\Delta s_n \sim n$ (See, the proof of Theorem 1 in Ref. [18] and the numerical result shown in Fig. 2(b) in Ref. [23].), it is found from Eq. (B.3) that $x_n$ is bounded by $C_1\sqrt{n\tau}<\left|x_n\right|<C_2n\tau$, where $C_1$ and $C_2$ are certain constants. Subsequently, we define the convergence rate of $\Delta s_n \rightarrow 0^+$ when $\tau \rightarrow 0^+$ as $r(\Delta s_n, \tau)$. For the above settings, we obtain the following Theorem. 
\begin{theorem}
Assuming the Euler scheme of the backward Loewner evolution described by Eqs. (B.3) and (B.4), the convergence rate $r(\Delta s_n, \tau)$ of $\Delta s_n \rightarrow 0^+$ as $\tau \rightarrow 0^+$ is approximated by
\begin{align}
r(\Delta s_n, \tau) &\sim O(\tau^{2}) \ \ \ \rm{for} \ \ \  \tau \le 1/ {\it N},\\
r(\Delta s_n, \tau) &\sim O(\tau^{3}) \ \ \ \rm{for} \ \ \  \tau > 1/ {\it N}.
\end{align}    
\end{theorem}
\begin{proof}
Let $\phi_1\left(n,\tau\right)$ and $\phi_2\left(n,\tau\right)$ be the non-negative functions defined as:
\begin{align}
\phi_1\left(n,\tau\right)&:=\frac{C_1n\tau^2+{n^2\tau^3}}{2n},\\ 
\phi_2\left(n,\tau\right)&:=\frac{C_2{n^2\tau^3}+{n^2\tau^3}}{2n}, 
\end{align} 
where $C_1$ and $C_2$ are certain positive constants.Using Eqs. (B.4), (B.7), and (B.8) and fixing $n$ as the end point of the time series $n=N$, we obtain the followings.    
\begin{align}
\phi_1\left(n=N,\tau\right)<\Delta s_N&=\frac{\tau (x^2+(N\tau)^2)}{2N\tau}<\phi_2\left(n=N,\tau\right), \ \ \rm{for} \ \ \  \tau > 1/{\it N}, \\ 
\phi_2\left(n=N, \tau\right)\leq\Delta s_N&=\frac{\tau (x^2+(N\tau)^2)}{2N\tau}<\phi_1\left(n=N,\tau\right), \ \ \rm{for} \ \ \ \tau \le 1/{\it N}. 
\end{align}
From Eqs. (B.9) and (B.10), we obtain Eqs. (B.5) and (B.6) as  $\tau \rightarrow 0^+$. 
\end{proof}
From the above result, the scaling of $r(\eta_s(n), \tau)$ is also derived as the following.
\begin{lemma}
The convergence rate $r(\eta_s(n), \tau)$ of $\Delta s_n \rightarrow 0^+$ as $\tau \rightarrow 0^+$ is approximated by
\begin{align}
r(\eta_s(n), \tau) &\sim O(\tau^{-0.25}) \ \ \ \rm{for} \ \ \  \tau \le 1/ {\it N},\\
r(\eta_s(n), \tau) &\sim O(\tau^{-0.5}) \ \ \ \ \rm{for} \ \ \  \tau > 1/ {\it N}.
\end{align}
\end{lemma}
\begin{proof}
From the relation $C_1\sqrt{n\tau}<\left|x_n\right|<C_2n\tau$, we obtain $r(x_n, \tau) \sim O(\tau^{1})$ for  $\tau  > 1/ {\it N}$ and $r(x_n, \tau) \sim O(\tau^{1/2})$ for  $\tau \le 1/ {\it N}$. Using this, the lemma follows from Theorem 2, Eq. (B.3) and Eq. (6).  
\end{proof}
We remark that the results in Fig.4 is consistent with Lemma B.2, as we choose $\tau=0.001$ and $N=10000$ in the numerical simulation in Sec. 4.


\begin{thebibliography}{0}

\bibitem{1} G. Carleo, I. Cirac, K. Cranmer, L. Daudet, M. Schuld, N. Tishby, et al. Machine learning and the physical sciences. {\it Rev. Mod. Phys} {\bf 91}, 045002 (2019), \url{https://doi.org/10.1103/RevModPhys.91.045002}.

\bibitem{2} Y. Tang, J. Kurths, W. Lin, E. Ott, and L. Kocarev, Introduction to Focus Issue: When machine learning meets complex systems: Networks, chaos, and nonlinear dynamics. {\it Chaos} {\bf 30}, 063151 (2020), \url{https://doi.org/10.1063/5.0016505}.  

\bibitem{3} F. Borra, A. Vulpiani, and M. Cencini, Effective models and predictability of chaotic multiscale systems via machine learning. {\it Phys. Rev. E} {\bf 102} 052203 (2020), \url{https://doi.org/10.1103/PhysRevE.102.052203}. 

\bibitem{4} L. Gammaitoni, and A. Vulpiani, Prediction and Inference: From Models and Data to Artificial Intelligence. {\it Found. Phys.} {\bf 54} 67 (2024), \url{https://doi.org/10.1007/s10701-024-00803-4}. 

\bibitem{5} J. J. Hopfield, Neural networks and physical systems with emergent collective computational abilities. {\it Proc. Natl. Acad. Sci.} {\bf 79} 2554 (1982), \url{https://doi.org/10.1073/pnas.79.8.2554}.

\bibitem{6} J. J. Hopfield, and D. W. Tank, “Neural” computation of decisions in optimization problems. {\it Biol. Cybern.}  {\bf 52} 141 (1985), \url{https://doi.org/10.1007/BF00339943}.

\bibitem{7} C.M. Bishop, and N.M. Nasrabadi, {\it Pattern recognition and machine learning} {Springer, New York, 2006}. 

\bibitem{8} J. Schmidhuber, Deep learning in neural networks: An overview. {\it Neural Netw.} {\bf 61}, 85 (2015), \url{https://doi.org/10.1016/j.neunet.2014.09.003}.

\bibitem{9} S. Goldt, and U. Seifert, Stochastic Thermodynamics of Learning. {\it Phys. Rev. Lett.} {\bf 118} 010601 (2017), \url{https://doi.org/10.1103/PhysRevLett.118.010601}. 

\bibitem{10} A. Seif, M. Hafezi, and C. Jarzynski, Machine learning the thermodynamic arrow of time. {\it Nat. Phys.} {\bf 17} 105 (2021), \url{https://doi.org/10.1038/s41567-020-1018-2}.

\bibitem{11} D.H. Wolpert, The stochastic thermodynamics of computation. {\it J. Phys. A}  {\bf 52}, 193001(2019), \url{https://doi.org/10.1088/1751-8121/ab0850}.

\bibitem{12} I. A. Gruzberg, and L.P. Kadanoff, The Loewner Equation: Maps and Shapes. {\it J. Stat. Phys.} {\bf 114} 1183 (2004), \url{https://doi.org/10.1023/B:JOSS.0000013973.40984.3b}. 

\bibitem{13} R.O. Bauer, {\it AFST: Math.} {\bf 12} 433 (2003).

\bibitem{14} T. Kennedy, Computing the Loewner Driving Process of Random Curves in the Half Plane. {\it J. Stat. Phys.} {\bf 131} 803 (2008), \url{https://doi.org/10.1007/s10955-008-9535-x}.

\bibitem{15} S. Rohde, and O. Schramm, Basic properties of SLE. {\it Ann. Math.} {\bf 161} 883 (2005), \url{https://doi.org/10.4007/annals.2005.161.883}. 

\bibitem{16} Y. Shibasaki, and M. Saito, Loewner driving force of the interface in the 2-dimensional Ising system as a chaotic dynamical system. {\it Chaos} {\bf 30} 113130 (2020), \url{https://doi.org/10.1063/5.0023261}.

\bibitem{17} Y. Shibasaki, Permutation-Loewner entropy analysis for 2-dimensional Ising system interface. {\it Physica A} {\bf 594} 126943 (2022), \url{https://doi.org/10.1016/j.physa.2022.126943}.

\bibitem{18} Y. Shibasaki, On the Role of Loewner Entropy in the Statistical Mechanics of the 2D Ising System. {\it PTEP} {\bf 2025} 023A02 (2025), \url{https://doi.org/10.1093/ptep/ptaf017}.

\bibitem{19} Y. Shibasaki, Loewner Theory for Stochastic Neuron Model. {\it Biophys. Rev. Lett.} {\bf 19} 183 (2024), \url{https://doi.org/10.1142/S1793048024500048}.

\bibitem{20} M. Kanagawa, P. Hennig, D. Sejdinovic, and B.K. Sriperumbudur, Gaussian Processes and Kernel Methods: A Review on Connections and Equivalences. {\it arXiv preprint} {arXiv:1807.02582} (2018). 

\bibitem{21} U.M.B. Marconi, A. Puglisi, L. Rondoni, and A. Vulpiani, Fluctuation–dissipation: Response theory in statistical physics. {\it Phys. Rep.} {\bf 461} 111 (2008), \url{https://doi.org/10.1016/j.physrep.2008.02.002}.

\bibitem{22} A. Sarracino, and A. Vulpiani, On the fluctuation-dissipation relation in non-equilibrium and non-Hamiltonian systems. {\it Chaos} {\bf 29} 083132 (2019), \url{https://doi.org/10.1063/1.5110262}. 

\bibitem{23} Y. Shibasaki, Fluctuation-dissipation theorem with Loewner time. {\it Europhys.Lett.} {\bf 139} 31001 (2022), \url{https://doi.org/10.1209/0295-5075/ac7b44}.

\bibitem{24} Y. Shibasaki, M. Saito, and K. Judai, Loewner time conversion for q-generalized stochastic dynamics. {\it J. Stat. Mech. Theor. Exp.} {\bf 2023} 083205 (2023), \url{https://doi.org/10.1088/1742-5468/acecfc}.

\bibitem{25} B. Lindner, Fluctuation-Dissipation Relations for Spiking Neurons. {\it Phys. Rev. Lett.} {\bf 129} 198101 (2022), \url{https://doi.org/10.1103/PhysRevLett.129.198101}. 

\bibitem{26} F. Puttkammer, B. Lindner, Fluctuation–response relations for integrate-and-fire models with an absolute refractory period. {\it Biol. Cybern.} {\bf 118} 7 (2024), \url{https://doi.org/10.1007/s00422-023-00982-9}. 

\bibitem{27} M. Costa, A.L. Goldberger, and C.K. Peng, Multiscale entropy analysis of biological signals. {\it Phys. Rev. E} {\bf 71} 021906 (2005), \url{https://doi.org/10.1103/PhysRevE.71.021906}. 

\bibitem{28} H.R. Maturana, and F.J. Varela, (2012). {\it Autopoiesis and cognition: The realization of the
living} {Reidel, Dordrecht, 1972} 

\bibitem{29} P.B. Andersen, The semiotics of autopoiesis. A catastrophe-theoretic approach. {\it Cybern. Hum. Know.} {\bf 2} 17 (1994). 

\bibitem{30} D. Eriksson, K. Dong, E. Lee, D. Bindel, A.G. Wilson, (2018). {\it Advance in neural information processing systems} {\bf 31}.

\bibitem{31} C. Beck, and F. Schögl, {\it Thermodynamics of chaotic systems: an introduction} {Cambridge University Press, 1993}. 

\bibitem{32} P. Billingsley, (1968) {\it Convergence of Probability Measures} {Wiley, New York, 1968}. 

\bibitem{33} P. Oikonomou, I. Rushkin, I.A. Gruzberg, and L.P. Kadanoff, Global properties of stochastic Loewner evolution driven by Lévy processes. (2008). {\it J. Stat. Mech. Theor. Exp.} {\bf 2008} P01019 (2008), \url{https://doi.org/10.1088/1742-5468/2008/01/P01019}.


\end{thebibliography}
\end{document}